\documentclass[10pt]{iopart}
\usepackage{epsfig,cite}
\def\lsim{\raise0.3ex\hbox{$<$\kern-0.75em\raise-1.1ex\hbox{$\sim$}}}
\def\gsim{\raise0.3ex\hbox{$>$\kern-0.75em\raise-1.1ex\hbox{$\sim$}}}

\def\bg#1{\mbox{\boldmath$#1$}}

\newcommand{\bec}{\begin{center}}
\newcommand{\enc}{\end{center}}
\newcommand{\beq}{\begin{equation}}
\newcommand{\eeq}{\end{equation}}
\newcommand{\ds}{\displaystyle}
\newcommand{\beqar}{\begin{eqnarray}}
\newcommand{\eeqar}{\end{eqnarray}}

\newcommand{\D}{{\cal D}}
\newcommand{\Pom}{{\hspace{ -0.1em}I\hspace{-0.25em}P}}

\newcommand{\MQQ}{M_{\scriptscriptstyle{Q\bar{Q}}}}
\begin{document}

\title{ Nuclear suppression at RHIC and LHC in Glauber-Gribov approach }
\author{
K~Tywoniuk\dag, I~Arsene\dag, L~V~Bravina\dag\ddag, A~B~Kaidalov\S, 
E~E~Zabrodin\dag\ddag
}
\address{\dag\
         Department of Physics, University of Oslo, Oslo, Norway}
\address{\ddag\
         Institute for Nuclear Physics, Moscow State University, 
         Moscow, Russia}
\address{\S
         Institute of Theoretical and Experimental Physics, 
         Moscow, Russia}

\begin{abstract}
The approach to problem of nuclear shadowing based on Gribov Reggeon
calculus is presented. Here the total cross section of $h A$ 
interaction is found in a parameter-free description, employing the
new data on the gluon density of the Pomeron, measured with high
precision at HERA, as input. The model is then applied for 
calculation of $J/\psi$ production in $d Au$ collisions at top RHIC
energy. It is shown that the theoretical estimates are in a very good
agreement with the PHENIX data, and further predictions for the 
$J/\psi$ suppression in $p Pb$ collisions at coming soon LHC are made.
\end{abstract}

\section{Introduction}
\label{sec1}

One of the ultimate goals of the heavy-ion collision programme at
ultra-relativistic energies is the search for fingerprints of a new 
state of matter, the so-called quark-gluon plasma (QGP). Since up to 
now no signals are observed which can be unambiguously attributed to 
the QGP formation in hadron$-$hadron ({\it hh\/}) or hadron$-$nucleus 
({\it hA\/}) interactions, such processes are used as 
reference ones for comparison with the nuclear ({\it A+A\/}) 
collisions. On the other hand, even the physics of ({\it hA\/}) 
interactions is not completely understood yet. For instance, 
experimentalists found a substantial decrease of the nuclear 
absorption in $J/\psi$ production in deuteron$-$gold ({\it dAu\/}) 
collisions at top RHIC energy, $\sqrt{s} = 200$\, AGeV, in comparison 
with proton$-$lead interactions at SPS, $E_{lab} = 158$\, AGeV 
\cite{phen_jpsi}. Here the dynamics of the collision is obviously 
changing with rising bombarding energy, and new effects come into 
play. To study these processes we employ the Gribov Reggeon theory
(GRT) \cite{gribov}, which enables one to take into consideration 
both soft and hard processes.

\begin{figure}[htb]
\begin{minipage}[t]{70mm}
\epsfig{file=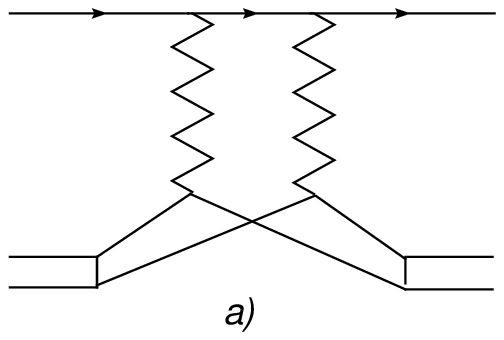,width=40mm}
\end{minipage}
\hspace{\fill}
\begin{minipage}[t]{70mm}
\epsfig{file=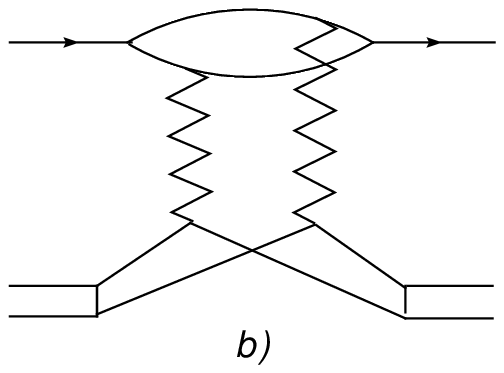,width=40mm}
\label{fig1}
\end{minipage}
\caption{(a) Planar and (b) non-planar diagrams of {\it hA} scattering.}
\end{figure}

At low incident energies the multiple scattering of incoming hadron on 
nuclei is described by the planar diagrams shown in Fig.~\ref{fig1}(a).
It appears that in this energy range the total cross section of the 
process, calculated within the GRT, is equal to that given by the
probabilistic Glauber model \cite{glaub}, which takes into account
successive elastic rescatterings of the incoming hadron on the 
nucleons of the target. 
Note, that the multiple scattering in the 
Glauber model leads to reduction of the total cross section. 
The situation changes dramatically at energies higher than a certain 
critical energy $E_{crit}$ \cite{gribov}. Above $E_{crit}$ the typical 
length of hadronic fluctuations 
becomes comparable or even exceeds the nuclear radius, thus leading to 
the coherent interaction of hadron constituents with several nucleons. 
The main contribution to the total cross section comes here from the
non-planar diagrams, depicted in Fig.~\ref{fig1}(b), while the 
contribution from the planar diagrams drops with rising energy 
inversely proportional to $E$. Since the interactions with different
nucleons of the target nucleus occur almost simultaneously, the 
space-time analogy to the Glauber rescattering series is lost. The
diffractive intermediate states should be taken into account
\cite{gribov}, and the total cross section is further reduced. The 
phenomenon is colloquially known as nuclear shadowing. It can be 
decomposed onto the quark shadowing and the gluon shadowing; the 
latter provides largest theoretical uncertainties.

\section{Inelastic shadowing in Gribov model}
\label{sec2}

\begin{figure}[htb]
\begin{minipage}[t]{65mm}
\bec
\epsfig{file=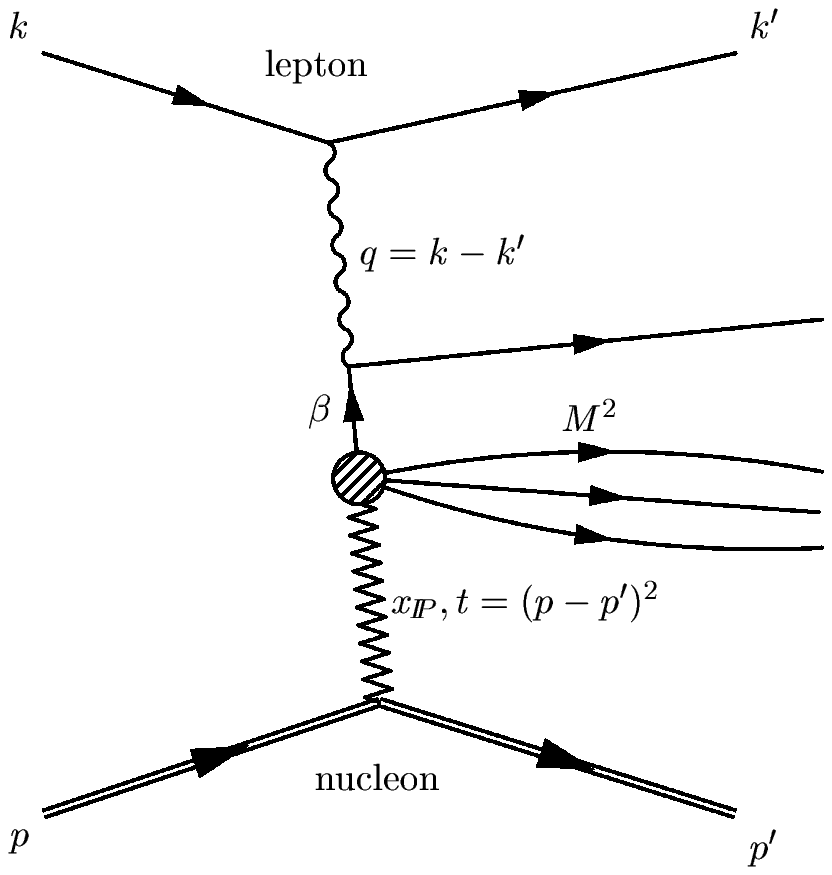,width=45mm}
\caption{
DDIS kinematical variables in the infinite momentum frame.
}
\label{fig2}
\enc
\end{minipage}
\hspace{\fill}
\begin{minipage}[t]{65mm}
\bec
\epsfig{file=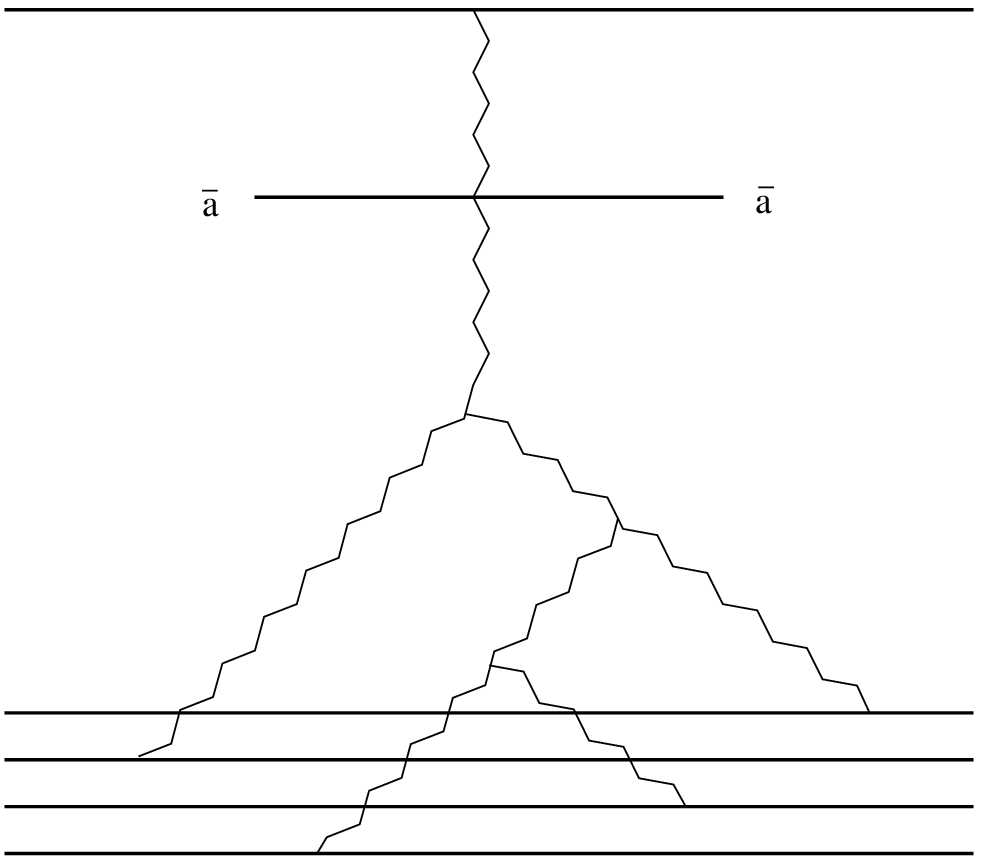,width=45mm}
\caption{
Enhanced diagram for the inclusive production of particle $a$.
}
\label{fig3}
\enc
\end{minipage}
\end{figure}

Below we will follow the procedure formulated in details in
\cite{Cap98,Arm03,Tyw07}. The {\it hA\/} cross
section is expanded in infinite series containing the contributions
from $1, 2 \ldots$ scatterings
\beq
\ds
\sigma_{hA}^{(tot)} = 
\sigma_{hA}^{(1)} + \sigma_{hA}^{(2)} + \ldots \ ,
\label{eq:sum}
\eeq
where the first term represents the sum of independent interactions
and the subsequent terms describe multiple inelastic interactions,
respectively, of the incoming probe with the
nucleons in the target nucleus. The contribution from the second
term, $\sigma_{h N}^{(2)}$, arises from the cut contribution
of the double rescattering diagrams. It appears that this term is
identical to minus the contribution from the diffractive cut, thus
linking the nuclear shadowing to the diffractive deep inelastic 
scattering (DDIS). The standard variables for the description of the
DDIS are $Q^2, x, M^2$ and $t$, or $x_\Pom$; they are depicted in 
Fig.~\ref{fig2}.
The variable $\beta = \frac{Q^2}{Q^2+M^2} = x /x_\Pom$ plays the same 
role for the Pomeron as the Bjorken variable, $x$, for the nucleon.
We have a reduction of the total $h A$ cross section by a term
\beqar
\ds
\label{eq:sec}
\sigma^{(2)}_{h A}& &\;=\; \; -4\pi A(A-1) \times \\
\nonumber
& & \;\int \mbox{d}^2b \, T_A^2 (b)
\int_{M^2_{min}}^{M^2_{max}} \mbox{d}M^2 \, \left[ \frac{\mbox{d}
    \sigma^{\D}_{h \scriptscriptstyle{N}} (Q^2, x_\Pom, \beta)}
{\mbox{d} M^2 \, \mbox{d}t}\right]_{t=0} \, F_A^2 (t_{min}) \; ,
\eeqar
where 
$$\ds T_A (b) = \int^{+\infty}_{-\infty} \mbox{d}z \, \rho_A ({\bg
  b}, z)$$ 
is the nuclear normalized density profile, $\int \mbox{d}^2b \,T_A (b) 
= 1$. The form factor $F_A$ is expressed as
\beq
\ds
\label{eq:FA}
F_A (t_{min}) \;=\; \int \mbox{d}^2b \, J_0 (\sqrt{-t_{min}}b) \, T_A
(b) \;,
\eeq
with $t_{min} = -m_N^2 x^2_\Pom$, and $J_0(x)$ denoting the Bessel
function of the first kind. 
Although 
Eq.~(\ref{eq:sec}) is an identity for nuclear densities which depend 
separately on ${\bf b}$ and $z$, we have checked \cite{jyu07} that 
calculations with an exact expression lead to negligible corrections. 
Since Eq.~(\ref{eq:sec}) is obtained under very general assumption, 
i.e. analyticity and unitarity, it can be applied for arbitrary 
values of $Q^2$ provided $x$ is very small. 
For a deuteron, the double 
rescattering contribution has the following form
\beq
\label{eq:deuteron}
\sigma_{h d}^{(2)} \;=\; -2 \int_{-\infty}^{t_{min}} \mbox{d}t \,
\int^{M^2_{max}}_{M^2_{min}} \mbox{d}M^2 \,
\frac{\mbox{d}\sigma^\D_{h N}}{\mbox{d}M^2 \mbox{d}t} \, F_D (t)
\;,
\eeq
where the deuteron form factor is roughly approximated as 
$F_D (t) = \exp (at)$, with $a = 40 \mbox{ GeV}^{-2}$ \cite{Cap98}.

The integration limit $M^2_{min}$ 
corresponds to the minimal mass of the diffractively produced hadronic 
system, $M^2_{min} = 4m_{\pi}^2 = 0.08 \mbox{GeV}^2$, and $M^2_{max}$ 
is chosen from the condition: $x_\Pom \le 0.1$.
It guarantees \cite{kaidalovrep} the disappearance of
nuclear shadowing at $x \sim 0.1$ in accord with experimental data.
Coherence effects are taken into account via $F_A (t_{min})$, which 
equals to 1 at $x \rightarrow 0$ and 
decreases with increasing $x$ due to the loss of coherence for 
$x > x_{crit} \sim \left( m_N R_A \right)^{-1}$. In the 
calculations a 3-parameter Woods-Saxon nuclear density profile with 
parameters from \cite{jager} has been used.

For higher order rescatterings we use the Schwimmer unitarization 
\cite{schwimmer} for the total $h A$ cross section which is 
obtained from a summation of fan-diagrams with triple-Pomeron 
interactions shown in Fig.~\ref{fig3}.  As was checked in 
\cite{Cap98}, this method provides results very close to other 
reasonable models, such as the quasi-eikonal model. The total cross 
section is then
\beq
\ds
\label{eq:sch}
\sigma_{h A}^{Sch} \;=\; \sigma_{h N} \, \int
\mbox{d}^2b \; \frac{A \, T_A (b)}{1 \,+\, (A-1) f(x, Q^2) T_A (b) } \;.
\eeq
Thus the total $h A$ cross section can be calculated
within the Glauber-Gribov model in a parameter-free way provided the 
total $h N$ cross section and the differential cross section for 
diffractive production are known.
To determine the shadowing for quarks and antiquarks one has to insert
in Eq.~(\ref{eq:sch})
\beq
\ds
\label{eq:f}
f(x, Q^2) \;=\; 4\pi\; \int_x^{x_\Pom^{max}} \mbox{d}x_\Pom \,
B(x_\Pom)\, \frac{F_{2 \D}^{(3)}(x_\Pom, Q^2, \beta)}{F_2 (x, Q^2)} 
\; F_A^2 (t_{min.}) \;.
\eeq
Here $F_2(x,Q^2)$ is the structure function for a nucleon,
$F_{2\D}^{(3)}(x_\Pom,Q^2,\beta)$ is the $t$-integrated diffractive 
structure function of the nucleon, 
$B(x_\Pom) = B_0 \,+\, \alpha_\Pom' \ln \frac{1}{x_\Pom}$, 
$\alpha_\Pom'$ is the slope of the Pomeron trajectory
$\alpha_\Pom (t) = \alpha_\Pom (0) + \alpha_\Pom' t$ (see 
\cite{H106}), and $F_A (t_{min})$ is given by Eq.~(\ref{eq:FA}). 
Note, that the real part of the diffractive amplitude in 
Eqs.~(\ref{eq:sec}),(\ref{eq:sch}),(\ref{eq:f}) is neglected. 
Similar expressions are valid for the gluon shadowing with 
substitutions $F_{2\D}^{(3)} \left(x_\Pom,Q^2,\beta \right) 
\rightarrow \beta g^\D (\beta,Q^2)$, $F_2 \left(x,Q^2 \right) 
\rightarrow x g\left(x,Q^2 \right)$, indicating gluon distributions 
in DDIS and in the proton, respectively.
Gluon distribution of the nucleon is taken from CTEQ6M 
parameterization \cite{CTEQ}.  We take information on the diffractive 
gluon distribution and Pomeron parameters from recent HERA 
measurements \cite{H106}. Further details of the developed approach 
can be found in \cite{Cap98,Arm03,Tyw07}. 

\section{Nuclear effects in heavy quarkonium production}
\label{sec3}

As was already mentioned in Introduction, the substantial decrease 
of the nuclear absorption in $J/\psi$ production, observed by PHENIX
collaboration in {\it dAu\/} collisions at RHIC \cite{phen_jpsi},
has attracted a lot of attention. Many models predict that absorptive
effects would increase or, at least, remain constant with rising
collision energy \cite{Braun98}. However, such a
behavior of absorptive cross section allows for a natural explanation 
in the framework of Gribov theory. We argue that the apparent 
observation of the reduction of $\sigma_{abs}^{J/\psi}$ can
be interpreted as a signal of the onset of coherent scattering
for heavy-state production.

The reason is as follows \cite{J_psi_07}.
At very high energies Abramovsky-Gribov-Kancheli (AGK) cutting rules 
\cite{AGK} lead to a cancellation of the Glauber-type 
diagrams in the central rapidity region, i.~e. for $x_F \approx 0$, 
and only enhanced diagrams (see Fig.~\ref{fig3}) 
\cite{Kancheli70} give non-zero contributions.
But both at non-zero $x_F$ and at lower energies the AGK
cancellation is not valid due to energy-momentum conservation 
\cite{Cap76}. This leads to an increase of effective ``absorption'' 
with rising $x_F$. The energy dependence of AGK violation is 
different for light- and heavy-state production because of the mass 
difference. For heavy quark states the mass $\MQQ$ of the heavy 
system introduces a new scale
\beq
\label{eq:scale}
s_M \;=\; \frac{\MQQ^2}{x_+} \, \frac{R_A m_N}{\sqrt{3}} \;,
\eeq
where $x_+ = \frac{1}{2}( \sqrt{x_F^2 + 4\MQQ^2/s} + x_F)$ is the
longitudinal momentum fraction of the heavy system. It was shown 
\cite{Boreskov91} that AGK cutting rules are changed at $s = s_M$. At 
energies below $s_M$ longitudinally ordered rescatterings of the heavy 
system take place. At $s > s_M$ the heavy state in the projectile
scatters coherently off the nucleons of a nucleus, and the 
conventional treatment of nuclear absorption is not adequate. In the 
central rapidity region the values of $s_M$ for $J/\psi$ are within 
the RHIC energy range. Accordingly, the effects of shadowing of
nuclear partons become important and can be calculated using Gribov
theory of nuclear structure functions in the region of $x_2 <
(m_N R_A)^{-1}$. Based on the above discussion we calculate the 
suppression of $J/\psi$ production in the central rapidity region in 
{\it pA\/} collisions at $\sqrt{s} = $ 200 GeV taking into account
shadowing effects \cite{jpsi_plb_07}. A similar approach, albeit with 
a simpler parameterization of nuclear shadowing, has also been 
considered in \cite{Cap06}.

PHENIX collaboration has measured the nuclear modification factor
(NMF) of $J/\psi$ production in {\it dAu\/} collisions at RHIC as a 
function of centrality \cite{phen_jpsi}. The centrality dependent 
NMF is defined as
\beq
\label{eq:nmf}
R_{dAu} \left( \langle N_{coll} \rangle \right) \;=\;
\frac{N_{inv}^{dAu} \left( \langle N_{coll} \rangle \right)}
{\langle N_{coll} \rangle \times N_{pp}^{inv}} \;,
\eeq
where the average number of nucleon-nucleon collisions $\langle
N_{coll} \rangle$ is obtained from the Glauber model for a given 
centrality.
\begin{figure}[htb]
\begin{minipage}[t]{60mm}
\epsfig{file=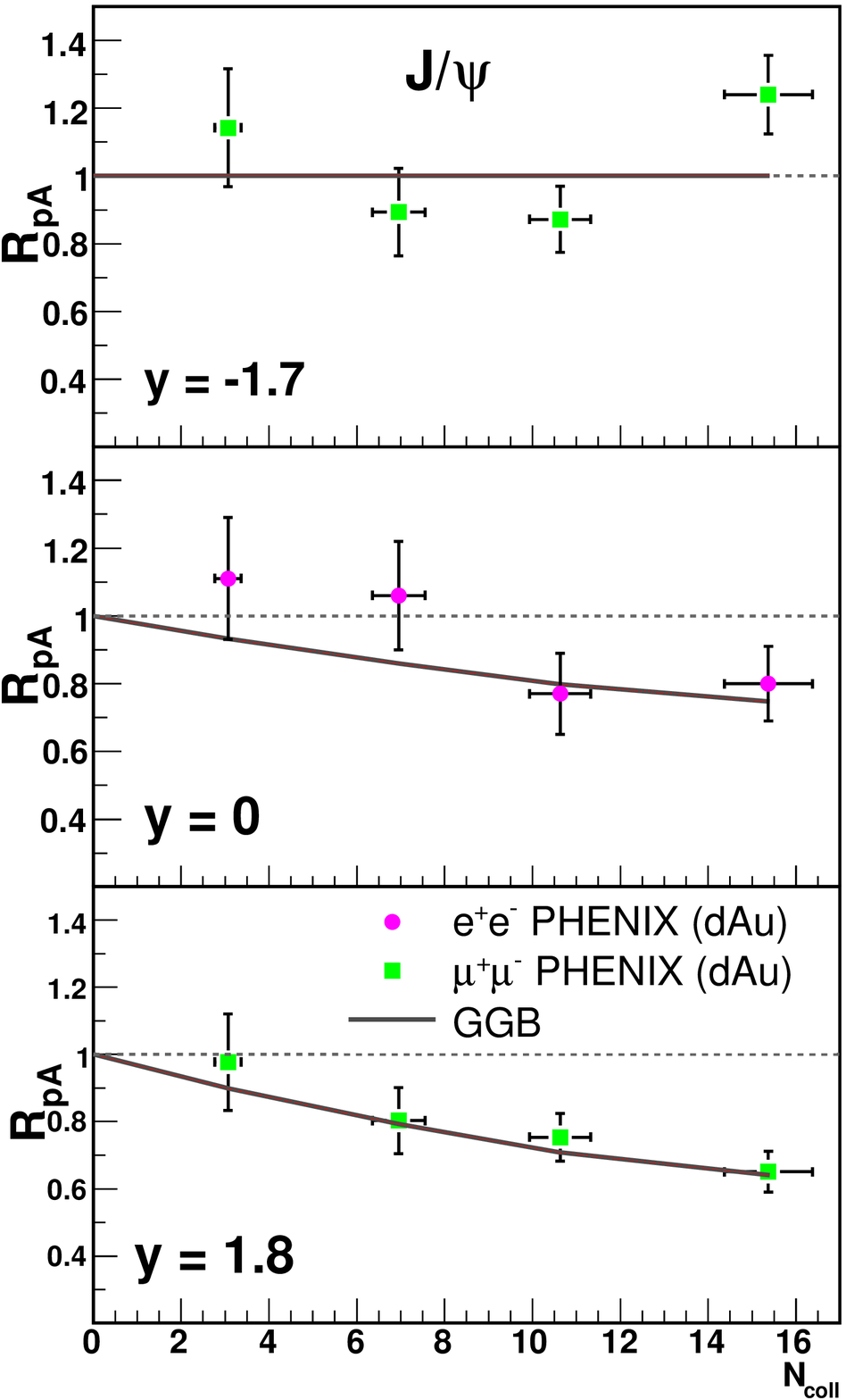,width=50mm}
\end{minipage}
\hspace{\fill}
\begin{minipage}[t]{70mm}
\epsfig{file=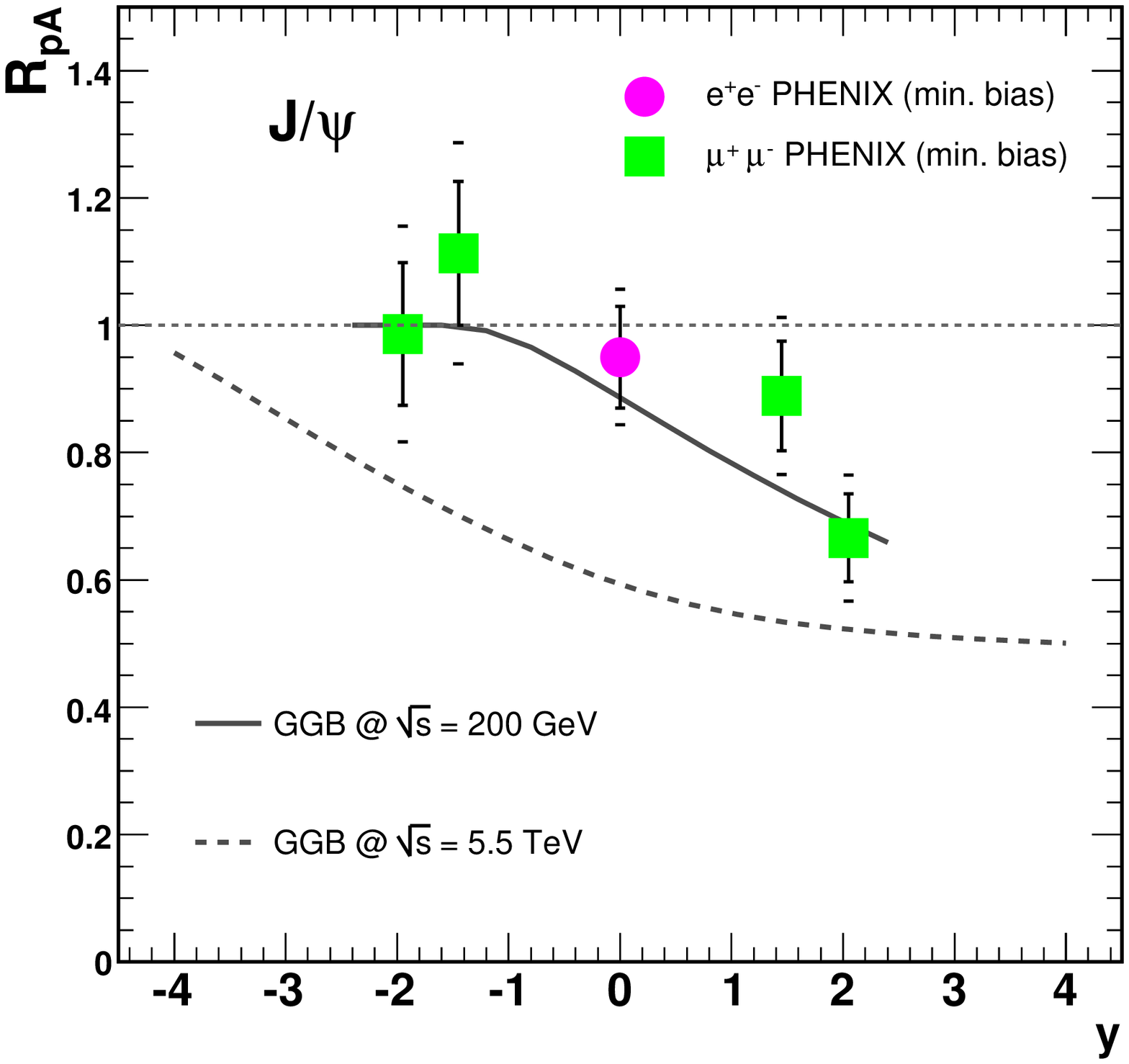,width=70mm}
\end{minipage}
\caption{
(a) Centrality and (b) rapidity dependence of the nuclear modification 
factor $R_{pA}^{J/\psi}$ in {\it d Au\/} and {\it p Pb\/} collisions 
at top RHIC and LHC energies. Data are taken from 
\protect\cite{phen_jpsi}.
}
\label{fig4}
\end{figure}

In the model considered here the suppression factor is given by
\beq
\ds
R_A(b,x,Q^2) \;=\; \frac{ A T_A(b)}{1 \,+\, (A-1)f(x,Q^2)T_A(b)}\ ,
\label{eq:R_A}
\eeq
where $f(x,Q^2)$ is given by Eq.~(\ref{eq:f}).
The summation of planar and non-planar diagrams provides us the 
following expression for the ratio of inclusive spectra in {\it dA\/}
and {\it pp\/} collisions
\beq
\ds
E \frac{\mbox{d}^3 \sigma_{dAu}}{\mbox{d}^3 p} (b) /
E \frac{\mbox{d}^3 \sigma_{pp}}{\mbox{d}^3 p}
\;=\; \int 
\mbox{d}^2s R_d( {\bf b},x_p)
R_{Au}({\bf s} - {\bf b},x_t) \, 
\eeq
where $x_{p (t)} = m_\perp \exp(\pm y)/\sqrt{s}$ 
(we take $m_\perp = Q$ and $Q = 4$ GeV).
This is what we compare to the nuclear modification factor presented 
by Eq.~(\ref{eq:nmf}).

The results of calculations are shown in
Fig.~\ref{fig4}(a) for the NMF at backward, mid- and forward 
rapidity. Since the model of gluon shadowing does not include 
anti-shadowing effects, the result is quite trivial in the backward 
hemisphere, although not inconsistent with the data. Anti-shadowing 
is assumed to be a 10$\%$ effect. At rapidity $y=0$ and $y=1.8$ the 
consistency with experimental data is quite good. 
Within other parameterizations of nuclear shadowing, the data has 
been shown to be consistent with an absorptive cross section of 1-2 mb 
\cite{phen_jpsi}, which is smaller than at lower energies. Whereas the
absorptive effects were predicted to increase with energy \cite{Bedj03}, 
our model provides a unique explanation of the competing effects of 
shadowing and absorption at high energies.
Description of 
nuclear modification factor $R_{dAu}$ for minimum bias {\it dAu\/} 
collisions at RHIC \cite{phen_jpsi} and predictions for {\it pPb\/} 
collisions at LHC are presented in Fig.~\ref{fig4}(b). While being 
accounted for about 10\% drop of the NMF at $y=0$ at RHIC, gluon 
shadowing at LHC should reduce the NMF of $J/\psi$ to $R_{pPb}^{J/\psi} 
\simeq 0.6$ at mid-rapidity and to $R_{pPb}^{J/\psi} \simeq 0.5$ at 
forward rapidities for minimum bias events.

\section{Conclusions}
\label{sec4}

We have discussed the vanishing of nuclear absorption and the
appearance of shadowing in the context of $J/\psi$ production at RHIC 
and LHC energies. Within the Gribov model this is related to a change 
of the space-time dynamics of the collision, going from incoherent, 
longitudinally ordered scattering at low energies to coherent 
scattering at higher energies. Within the same model we calculated 
gluon shadowing using recent measurements of gluon diffractive parton 
density at HERA as input.

A comparison to experimental data from RHIC at mid-rapidity and away 
from it shows good agreement. We predict a stronger shadowing effect at
LHC. Gluon shadowing is also a crucial input for the correct treatment
of final-state effects in nucleus-nucleus collisions.

{\bf Acknowledgments.} The authors are grateful to N. Armesto,
K. Boreskov, A. Capella and C. Pajares for interesting discussions
and valuable comments.
This work was supported by the Norwegian Research Council (NFR) under 
contract No. 166727/V30, RFBF-06-02-17912, RFBF-06-02-72041-MNTI, 
INTAS 05-103-7515, grant of leading scientific schools 845.2006.2 and 
support of Federal Agency on Atomic Energy of Russia.


\section*{References}
\begin{thebibliography}{99}

\bibitem{phen_jpsi} 
PHENIX Collaboration, Adler~S~S et al. 2003 \PRL {\bf 96} 012304

\bibitem{gribov} 
Gribov~V~N 1969 {\it Sov. Phys.--JETP\/} {\bf 29} 483; 
1968 {\it Sov. Phys.--JETP\/} {\bf 26} 414;
1970 {\it Sov. Phys.--JETP\/} {\bf 30} 709

\bibitem{glaub} 
Glauber~R~J 1959 {\it Lectures in Theoretical Physics\/}
{\bf 1} (ed. by Brittin~W~E and Dunham~L~G) 315

\bibitem{Cap98} 
Capella~A, Kaidalov~A~B, Merino~C, Perterman~D, Tran Thanh Van~J 
1998 {\it Eur. Phys. J.\/} {\bf C5} 111

\bibitem{Arm03} 
Armesto~N, Capella~A, Kaidalov~A~B, L\'opez-Albacete~J, Salgado~C~A
2003 {\it Eur. Phys. J.\/} {\bf C29} 531

\bibitem{Tyw07} 
Tywoniuk~K, Arsene~I, Bravina~L, Kaidalov~A~B, Zabrodin~E
2007 \PL {\bf B657} 170 (\texttt{arXiv:0705.1596[hep-ph]})

\bibitem{jyu07} 
Tywoniuk~K, Arsene~I, Bravina~L, Kaidalov~A~B, Zabrodin~E
2007 \texttt{arXiv:0709.1582[hep-ph]}

\bibitem{kaidalovrep} 
Kaidalov~A~B 1979 {\it Phys. Rep.\/} {\bf 50} 157

\bibitem{jager} 
De~Jager~C~W, De~Vries~H, De~Vries~C 1974
{\it Atom.~Data~Nucl. Data Table\/} {\bf 14} 479

\bibitem{schwimmer} Schwimmer~A 1975 \NP {\bf B94} 445

\bibitem{H106}
H1 Collaboration, Aktas~A et al. 2006 
{\it Eur. Phys. J.\/} {\bf C48} 715; {\it Eur. Phys. J.\/} {\bf C48} 749

\bibitem{CTEQ}
Pumplin~J, Stump~D~R, Huston~J, Lai~H~L, Nadolsky~P, Tung~W~K 
2002 {\it JHEP\/} {\bf 0207} 012

\bibitem{Bedj03} M.~Bedjidian {\it et al.} 2003 
\texttt{arXiv:hep-ph/0311048}

\bibitem{Braun98} 
Braun~M~A, Pajares~C, Salgado~C~A, Armesto~N, Capella~A,
1998 \NP {\bf B509} 357
\nonum
Kopeliovich~B, Tarasov~A, Hufner~J 2001 \NP {\bf A696} 669

\bibitem{J_psi_07} 
Tywoniuk~K, Arsene~I, Bravina~L, Kaidalov~A~B, Zabrodin~E
2007 \texttt{arXiv:0708.3801[hep-ph]}

\bibitem{AGK} 
Abramovsky~A~V, Gribov~V~N, Kancheli~O~V
1974 {\it Sov. J. Nucl. Phys.\/} {\bf 18} 308

\bibitem{Kancheli70} 
Kancheli~O~V 1970 {\it JETP Lett.\/} {\bf 11} 267
\nonum
Mueller~A~H 1970 \PR {\bf D2} 2963

\bibitem{Cap76} 
Capella~A, Kaidalov~A~B 1976 \NP {\bf B111} 477

\bibitem{Boreskov91} 
Boreskov~K~G, Capella~A, Kaidalov~A~B, Tran~Thanh~Van~J
1993 \PR {\bf D47} 919

\bibitem{jpsi_plb_07} 
Arsene~I, Bravina~L, Kaidalov~A~B, Tywoniuk~K, Zabrodin~E
2007 \texttt{arXiv:0711.4672[hep-ph]}

\bibitem{Cap06} 
Capella~A, Ferreiro~E 2006 \texttt{arXiv:hep-ph/0610313}

\end{thebibliography}
\end{document}